\documentclass[twocolumn,showpacs,preprintnumbers,amsmath,amssymb]{revtex4}

\usepackage{graphicx}% Include figure files
\usepackage{dcolumn}% Align table columns on decimal point
\usepackage{bm}% bold math

\begin{document}

\preprint{APS/123-QED}

%\title{Manuscript Title:\\with Forced Linebreak}% Force line breaks with \\
\title{Directed deterministic classical transport: symmetry breaking and beyond}
% Force line breaks with \\

%\author{Lucia Cavallasca}
\author{Lucia Cavallasca}%, Roberto Artuso and Giulio Casati}%
\email{lucia.cavallasca@uninsubria.it}
\author{Roberto Artuso}%
\author{Giulio Casati}%

\affiliation{Center for Nonlinear and Complex Systems and Dipartimento di Fisica e Matematica, \\
Universit\`a dell'Insubria, Via Valleggio 11, 22100 Como, Italy \\ CNISM,
Unit\`{a} di Como, Via Valleggio 11, 22100 Como, Italy \\
Istituto Nazionale di Fisica Nucleare, Sezione di Milano,
Via Celoria 16, 20133 Milano, Italy}

%\affiliation{

\date{\today}% It is always \today, today,

\begin{abstract}
We consider transport properties of a double $\delta$--kicked system, in a regime where all the symmetries (spatial and temporal) that could prevent directed transport are removed. We analytically investigate the (non trivial) behavior of the classical current and diffusion properties and show that the results are in good agreement with numerical computations. The role of dissipation for a meaningful classical ratchet behavior is also discussed.
%secondary publications and for information retrieval purposes. Valid
%PACS numbers may be entered using the \verb+\pacs{#1}+ command.
\end{abstract}

\maketitle

\section{\label{sec:level1}Introduction\protect\\ }
Dynamical systems exhibit an extremely rich variety of behavior with regards to transport properties. For instance, it has been known since a number of years, that the chaotic nature of the dynamics may induce stochastic-like properties in a deterministic system, like normal diffusion, as it happens for a random walk \cite{diff}. A full understanding of how dynamics precisely determines the nature of transport in unbounded systems is however still not fully accomplished (for instance deterministic transport is observed also for non-chaotic, yet mixing, systems \cite{dnon}): moreover there is a wide set of systems for which such properties are quite subtle (for instance anomalous behavior may appear in systems with a mixed phase space as a result of long sticking times of chaotic trajectories in the vicinity or regular islands \cite{andiff}).

A property which has recently attracted much attention is the ``ratchet effect", namely the generation of transport with a preferred direction, in systems without a net driving force (or even against a small applied bias) \cite{ratrev}. While stochastic ratchets are rather well understood, we are here interested in a purely deterministic setting, the starting point being in our case a family of area-preserving maps on a cylindrical (infinite) phase space. As remarked in \cite{kE} this generally requires a proper definition of what is meant by ``ratchet behavior", as directed transport is easily achieved even for a free system, once the starting velocity is different from zero. A meaningful notion of ratchet behavior is that of getting a non-zero current for generic initial conditions without the action of a net force, provided fluctuations of the current are not so wide to overwhelm the effect. Such a behavior has been thoroughly investigated in \cite{kE}, for systems with a compact phase space: a sum rule has been established, allowing for a precise theoretical estimate of the ratchet current. 

We address however a completely different situation, where unbounded transport is not a priori ruled out by confining invariant structures, and the expected behavior is normal diffusion, with null average momentum. This choice is motivated both as it is theoretically challenging, in establishing examples of fully underdamped ratchets, and as  modern cold atoms physics makes such systems good candidates for real experiments \cite{tania1,tania2,copi}. Though the system under investigation may be easily quantized (see \cite{copi,2harmd,Tq}) we here address the classical setting only: our results will provide evidence that transport properties, once we break up space-time symmetries are quite non trivial, but a classical good candidate for ratchet behavior requires an additional ingredient, which will be identified with dissipation in the present example.
\subsection{\label{sec:level2}The model}
The paradigmatic example of transport in low dimensional Hamiltonian dynamics is provided by the Chirikov-Taylor standard map
\begin{eqnarray} 
p_{n+1}\,&=&\, p_n+k \sin (\theta_n ) \label{st-map} \\
\theta_{n+1}\,&=&\,\theta_n + p_{n+1} \nonumber
%\label{st-map}
\end{eqnarray}
living in a cylindrical phase space $(\theta,p) \in \mathbb{S}^1\times\mathbb{R}$. Such a map does not only provide a case study to explore dynamical scenarios that arise as the nonlinear parameter $k$ is varied, but it also allows to consider novel features that appear upon quantization, most notably the so-called quantum dynamical localization \cite{loc}. Such a map exhibit symmetric transport properties: namely no average current ($\langle p_t-p_0 \rangle=0$) and linear growth of the variance (normal diffusion) in the regime $k>>1$ \cite{sm-anom}.

As observed in the seminal paper \cite{Flach}, searching for directed transport involves as a starting point a modification of the map (\ref{st-map}) in such a way to break time-space reversal symmetry. This can be accomplished in a number of ways \cite{tania1,tania2,copi}; we choose to follow the suggestion in \cite{copi}, as it is both theoretically simple and experimentally clean: namely unevenly spaced, phase shifted, kicks are introduced, described by a potential of the form 
\begin{eqnarray}
&V_{\phi, \xi}(\theta,t)=&\\  \nonumber
\\ \nonumber
&k\cos{(\theta)}\sum_{n=-\infty}^{\infty}\delta(t-n\tau)+&\\ \nonumber
\\ \nonumber
&k\cos{(\theta-\phi)}\sum_{n=-\infty}^{\infty}\delta(t-n\tau-1+\xi).&
\end{eqnarray}
The corresponding Hamiltonian is
\begin{eqnarray}
&H_{\phi, \xi}(p,\theta,t)=\frac{1}{2}p^2+V_{\phi, \xi}(\theta,t).&
\end{eqnarray}
$p$ and $\theta$ being conjugate variables. The corresponding discrete dynamics over a full period (corresponding to a pair of kicks) is written as
\begin{equation}
\left\{
\begin{array}{lcl}
p_{n+1}&=& p_{n} +k\sin(\theta_{n})+\\ 
\\
&&k\sin(\theta_{n}+\xi p_{n} +\xi k\sin(\theta_{n})-\phi)\\
\\
\theta_{n+1}&=&\theta_{n}+p_n+k\sin(\theta_{n})+\\
\\
&&(1-\xi) k\sin(\theta_{n}+\xi p_{n} +\xi k\sin(\theta_{n})-\phi).
\end{array}
\right. 
\label{map}
\end{equation}
We immediately point out an important feature: the extended map (\ref{map}) has an associated torus map, of size $2 \pi M$ in momentum, only for $\xi=N/M$, namely
for commensurate kicking times. 

Our primary goal will be to investigate transport properties of the map (\ref{map}), in particular the first two moments $\langle p_t -p_0 \rangle$ and $\langle (p_t-p_0)^2 \rangle$, especially for a choice of parameters' values breaking space-time symmetries: in the present case this is accomplished by choosing $\xi \neq \left\{ 0, \, 1/2 \right\} $ and $\phi \neq \left\{ 0,\, \pi \right\}$. 
\subsection{\label{sec:level3}Transport}
\begin{figure}[!htb]
\begin{center}
\includegraphics[height=6.3cm]{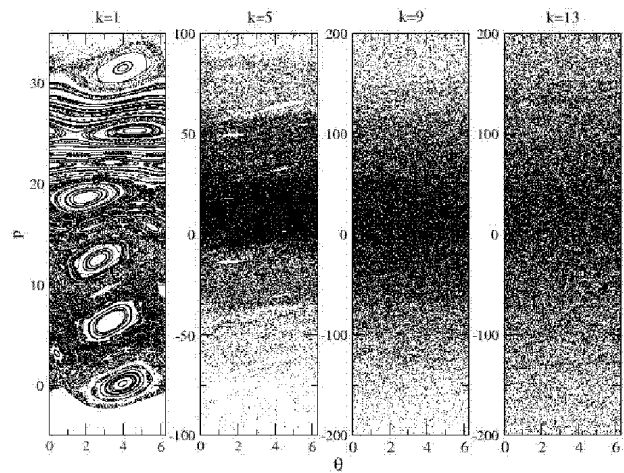}
\end{center}
\caption{\small{Phase space surface of section. $\phi=\pi/2$, $\xi=0.2$ and five different (increasing) values for the kick strength $k$. 200 iterates; 300 initial conditions in $(0,2\pi)\times(0,2\pi/\xi)$.}}
\label{fig1}
\end{figure}

\begin{figure}[!htb]
\begin{center}
\includegraphics[height=6.0cm]{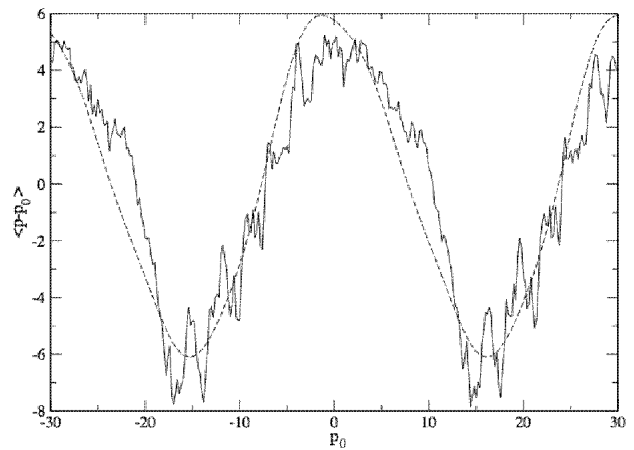}
\end{center}
\caption{\small{$\langle p-p_{0}\rangle$ after 50 couples of kicks, averaged over 100000 initial conditions, versus $p_{0}$. $k=9$, $\xi=0.2$, $\phi=\pi/2$. Full line: numerical results; dashed line: analytic estimate.}}
\label{fig2}
\end{figure}

\begin{figure}[!htb]
\begin{center}
\includegraphics[height=6.0cm]{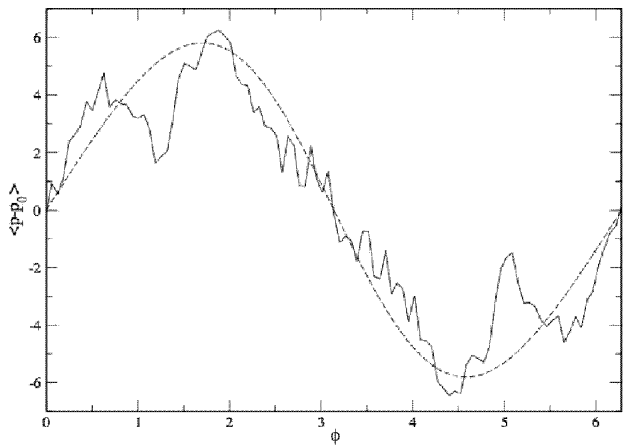}
\end{center}
\caption{\small{$\langle p-p_{0}\rangle $ after 50 couples of kicks, averaged over 100000 initial conditions, versus $\phi$. $k=9$, $\xi=0.2$, $p_{0}=0$. Full line: numerical results; dashed line: analytic estimate.}}
\label{fig3}
\end{figure}

\begin{figure}[!htb]
\begin{center}
\includegraphics[height=6.0cm]{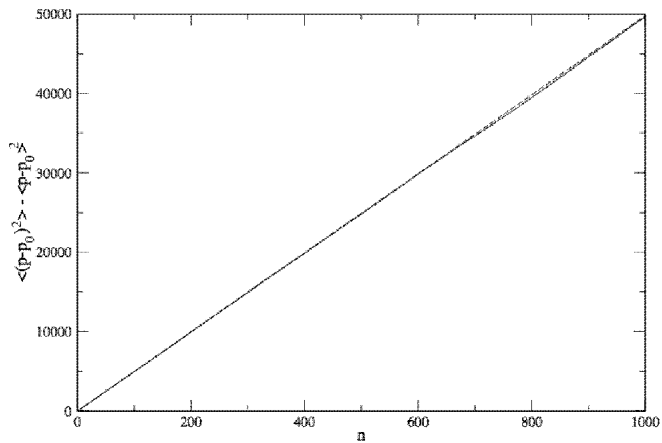}
\end{center}
\caption{\small{$\langle (p-p_{0})^{2}\rangle-\langle p-p_{0} \rangle^{2}$, averaged over 100000 initial conditions, versus time (expressed in kicks couples). $k=9$, $\xi=0.2$, $\phi=\pi/2$, $p_{0}=0$. Full line: numerical results; dashed line: analytic estimate.}}
\label{fig4}
\end{figure}

\begin{figure}[!htb]
\begin{center}
\includegraphics[height=6.0cm]{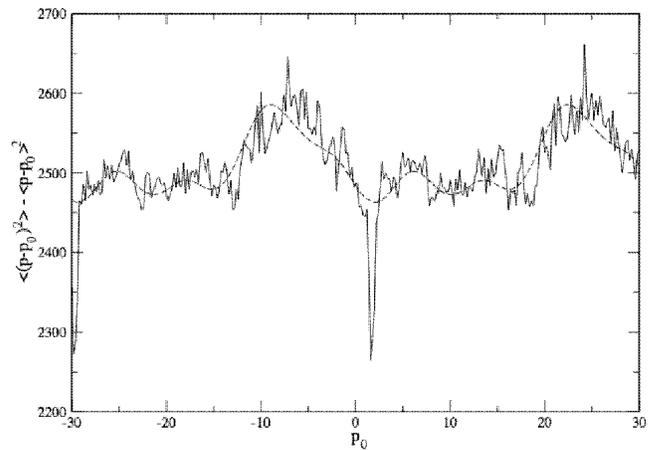}
\end{center}
\caption{\small{$\langle (p-p_{0})^{2}\rangle-\langle p-p_{0} \rangle^{2}$ after 50 couples of kicks, averaged over 100000 initial conditions, versus $p_{0}$. $k=9$, $\xi=0.2$, $\phi=\pi/2$. Full line: numerical results; dashed line: analytic estimate.}}
\label{fig5}
\end{figure}

First of all we observe that if we fix a pair $\xi,\,\phi$ and vary the nonlinear parameter $k$, a standard map-like scenario appears (see fig. (\ref{fig1})): namely for small $k$ KAM invariant structures create a barrier to unbounded transport, but these are destroyed for larger nonlinearity, allowing in principle unbounded trajectories on the cylinder. 
%\\
%\texttt{F1 -->  1}
\\
\\
Though transport may be \textit{anomalous} if sticking regularity regions influence motion in the chaotic sea, the typical behavior is normal diffusion (linear growth of the variance), while symmetry breaking generally leads to a nonzero first moment. To be more precise we introduce the following notation: $\Pi_{\chi_0}^{1,2}(t)$ will denote the first two moments at time $t$, obtained by averaging over a set $\chi_0$ of initial conditions: in particular a natural choice is to consider initial sets ${\cal M}_{p_0}$ ($p_0$ fixed, $\theta_0$ uniformly distributed on $[0,\,2\pi)$). For a wide choice of parameter values we observe that, after a transient, $\Pi_{{\cal M}_{p_0}}^{1}(t)$ saturates to an asymptotic value $\tilde{\Pi}_{p_0}^{1}$, which has a nontrivial dependence on $p_0$ (see the full curve in fig. (\ref{fig2})). 
%\\
%\texttt{F2--> 9}
\\
\\
As pointed out in \cite{copi} in this case we may obtain current reversal by tuning the phase shift between kicks; if we start from ${\cal M}_0$, as a matter of fact, $\tilde{\Pi}_{0}^{1}$ changes sign if we go from $\phi=\tilde{\phi}$ to $\phi=-\tilde{\phi}$ (such a property is also consistent with our analytic estimates, as we will see in the next section), see the full line in fig. (\ref{fig3}). 
%\\
%\texttt{F3--> 11}
\\
\\
So, by an appropriate choice of initial conditions we get an asymptotic momentum different from zero, and current reversal is easily obtained by tuning the phase shift between pairs of kicks; this can hardly be termed a ratchet behavior as the momentum distribution is broad, with a diffusive spread (see fig. (\ref{fig4})). 
%\\
%\texttt{F4--> 22}
\\
\\
Notice that also the diffusion constant exhibits dependence upon the starting set ${\cal M}_{p_0}$ as shown by the full line in fig. (\ref{fig5}).
%\\
%\texttt{F5--> 18}
\\
\\

\begin{figure}[!htb]
\begin{center}
\includegraphics[height=6.3cm]{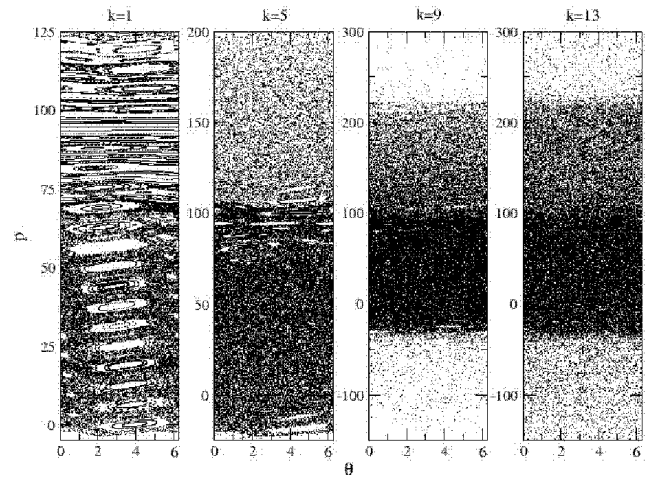}
\end{center}
\caption{\small{Phase space surface of section. $\phi=\pi/2$, $\xi=0.05$ and five different values for the kick strength $k$. 200 iterates; 300 initial conditions in $(0,2\pi)\times(0,2\pi/\xi)$.}}
\label{fig6}
\end{figure}

As observed in \cite{tania2} if the two kicks take place at close times peculiar effects may arise, in the sense that a cellular structure of the phase space emerges (see fig. (\ref{fig6})). 
%\\
%\texttt{F6--> 2}
\\
\\
More precisely this feature originates by requiring that the product $k \cdot \xi$ is small: to keep $k\xi$ small, we cannot lower $k$ too much, otherwise the phase space is no more fully chaotic. The cells are of size $2\pi/\xi$, separated by momenta $p=\pm (2m+1)\pi/\xi+\phi/\xi$, where $m=0,1,2...$; indeed at these momentum values (called trapping momentum) $p_{n+1}\simeq p_{n}$. They are not symmetric with respect to $p=0$, reflecting the fact that, with broken temporal and spatial symmetry, we expect a net current; if $\phi$ is equal to zero or $\pi$ the cells are symmetric with respect to $p=0$ (no net current without breaking the spatial symmetry).
The darkest cell corresponds to the cell where initial conditions are located.

\begin{figure}[!htb]
\begin{center}
\includegraphics[height=6.0cm]{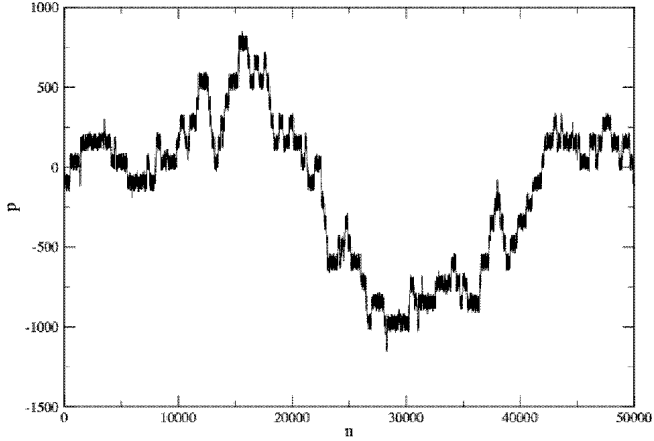}
\end{center}
\caption{\small{A typical trajectory (momentum versus time, expressed in kicks couples); $k=13$, $\xi=0.05$, $\phi=\pi/2$.}}
\label{fig7}
\end{figure}

Connected to this structure of the phase space, a typical trajectory spends a lot of time trapped in a cell before escaping onto another one, see figure (\ref{fig7}).
%\\
%\texttt{F7--> 3}
\\
\\

\begin{figure}[!htb]
\begin{center}
\includegraphics[height=6.0cm]{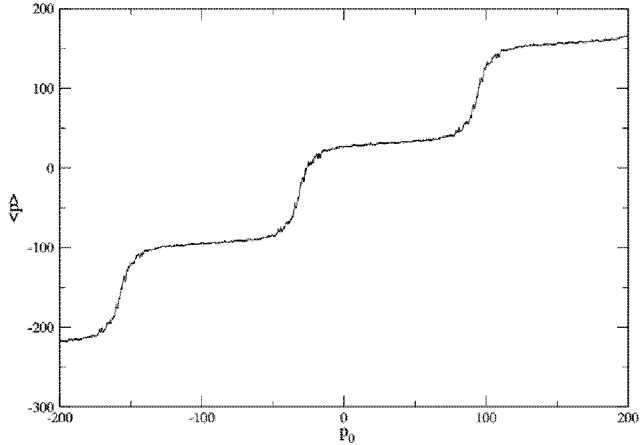}
\end{center}
\caption{\small{$\langle p\rangle $ after 50 couples of kicks, averaged over 10000 initial conditions, versus $p_{0}$. $k=13$, $\xi=0.05$, $\phi=\pi/2$.}}
\label{fig8}
\end{figure}

Transport is strongly dependent on initial conditions, in particular different behaviors emerge from initial conditions inside momentum cells or at cells boundaries, see for instance fig. (\ref{fig8}), where the asymptotic value of momentum is plotted.
%\\
%\texttt{F8--> 8}
\\
\\

Here cell trapping makes the dependence upon initial data more transparent: unless we go too close to the cell boundary, averages don't depend on the exact initial condition but essentially only upon which cell we start on. % (unless we go too close to the boundary): the average momentum exhibit some ``ergodic"-like behavior \cite{ML}. 
As remarked in \cite{tania2}, which we follow closely, near the cells borders the map can be approximated by the standard map (with a suitable kick strength). Let us introduce the rescaled variable $p^{(\xi)}=\xi p$ and the effective kick strength $k^{(\xi)}=\xi k$.

We choose an initial condition near the trapping momentum: $p_{0}^{(\xi)}=(2m+1)\pi+\phi+\delta p$ and $m=0$. We get:
\begin{equation}\nonumber
\begin{array}{rcl}
p_{1}^{(\xi)}&=&p_{0}^{(\xi)}+k^{(\xi)}\sin(\theta_{0})+\\
&&\\
                     &  &k^{(\xi)}\sin(\theta_{0}+p_{0}^{(\xi)}+k^{(\xi)}\sin(\theta_{0})-\phi)\\
%&=&p_{0}^{\xi}+k^{\xi}\sin(\theta_{0})+k^{\xi}\sin(\theta_{0}+\pi+\phi+\delta %p+k^{\xi}\sin(\theta_{0})-\phi)\\
%&=&p_{0}^{\xi}+k^{\xi}\sin(\theta_{0})+k^{\xi}[\sin(\theta_{0}+\pi)\cos(\delta %p+k^{\xi}\sin(\theta_{0})+\cos(\theta_{0}+\pi)\sin(\delta p+k^{\xi}\sin(\theta_{0})]\\
%&\simeq&p_{0}^{\xi}+k^{\xi}\sin(\theta_{0})-k^{\xi}\sin(\theta_{0})1-k%^{\xi}\cos(\theta_{0})(\delta p+k^{\xi}\sin(\theta_{0})\\
&&\\
&\simeq&p_{0}^{(\xi)}-(k^{(\xi)})^{2}\cos(\theta_{0})\sin(\theta_{0})-k^{(\xi)}\delta p\cos(\theta_{0}).
\end{array}
\end{equation}
Exactly in the middle of the trapping region $\delta p=0$ \cite{cosin}:
\begin{equation}\nonumber
p_{1}^{(\xi)}\simeq p_{0}^{(\xi)}-\frac{(k^{(\xi)})^{2}}{2}\sin(2\theta_{0})
\end{equation}
or, if we turn back to the original variables:
\begin{equation}\nonumber
p_{1}\simeq p_{0}-\frac{k^{2}}{2}\xi \sin(2\theta_{0})
\end{equation}

In a similar way for the angle variable:
\begin{equation}\nonumber
\begin{array}{rcl}
\theta_{1}&=&\theta_{0}+p_{0}+k\sin(\theta_{0})+\\
&&\\
&&(1-\xi)k\sin(\theta_{0}+p_{0}^{(\xi)}+k^{(\xi)}\sin(\theta_{0})-\phi)\\
%&\simeq&\theta_{0}+p_{0}+k\sin(\theta_{0})+(1-\xi)k(-\sin(\theta_{0})-%\cos(\theta_{0})(\delta p+k^{\xi}\sin(\theta_{0}))\\
&&\\
&\simeq&\theta_{0}+p_{0}+\xi k\sin(\theta_{0})-\\
&&\\
&&(1-\xi)k\cos(\theta_{0})(\delta p+k^{(\xi)}\sin(\theta_{0}))
\end{array}
\end{equation}
that, when $\delta p=0$, can be approximated to
\begin{equation}\nonumber
\theta_{1}=\theta_{0}+p_{2}
\end{equation}
%(we used: $k^{2}\xi\rangle \rangle k^{2}\xi^{2},k\xi$).
%\\
%\texttt{F9--> 1 from map equations}
\\
\begin{figure}[!htb]
\begin{center}
\includegraphics[height=6.3cm]{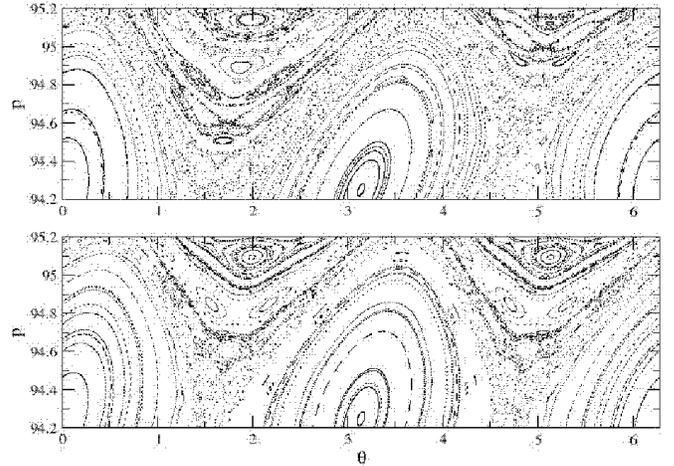}
\end{center}
\caption{\small{Local phase space. $k=4$, $\xi=0.05$, $\phi=\pi/2$ (see text).}}
\label{fig9}
\end{figure}
\\
In figure (\ref{fig9}) we compare the local phase space around the momentum cell border in the double-kicked system (top picture) and  in a $\sin(2x)$ single-kicked standard map with nonlinear parameter $K=k^{2}\xi/2$ (bottom picture).

\section{Estimates for the moments}
Analytic estimates for the first and the second moment of momentum distribution may be obtained by a standard technique \cite{PSFsm}, consisting in a Fourier expansion of the $\delta$-deterministic propagator, followed by inspection of leading contributions. If we start from an initial distribution $P(p,\theta,0)$ we may write the average of the $M$-th moment as
\begin{eqnarray}
&\langle (p_{N}-p_{0})^{M}\rangle =&\\
\nonumber\\
&\int Q(\theta_{N},p_{N},t_{N}|\theta,p,0)P(\theta,p,0) (p_{N}-p)^{M}d\theta_{N} d\theta dp_{N} dp&\nonumber
\label{app1}
\end{eqnarray}
where $Q(\theta_{N},p_{N},t_{N}|\theta,p,0)$ is the conditional probability of having $\theta_{N},p_{N}$ at time $t_{N}$ given that at time $0$ we have $\theta,p$. In particular averages corresponding to ${\cal M}_{p_0}$ have an initial distribution
\begin{equation}
P(\theta,p,0)=\frac{1}{2\pi}\delta(p-p_{0}).
\label{app2
}\end{equation}
$Q$ obeys the recursion property
\begin{eqnarray}
&Q(\theta_{N},p_{N},t_{N}|\theta_{0},p_{0},0)=&\\
\nonumber\\
&\int Q(\theta_{N},p_{N},t_{N}|\theta_{i},p_{i},t_{i})Q(\theta_{i},p_{i},t_{i}|\theta_{0},p_{0},0)d\theta_{i}dp_{i}&\nonumber
\label{app3}
\end{eqnarray}
and may be expressed in terms of the deterministic $\delta$ propagators:
\begin{eqnarray}
&Q(\theta_{i},p_{i},t_{i}|\theta_{i-1},p_{i-1},t_{i-1})=&\\
\nonumber\\
&\sum_{k_{i}=-\infty}^{+\infty}\delta(p_{i}-p_{i-1}+V'(\theta_{i-1}))&\nonumber\\
\nonumber\\
&\delta(\theta_{i}-\theta_{i-1}-(t_{i}-t_{i-1})(p_{i-1}-V'(\theta_{i-1}))+2\pi k_{i})\nonumber
\label{app4}
\end{eqnarray}
where the sum over $k_{i}$ occurs because $\theta$ has the angular topology. 

Also $(p_{N}-p_{0})$ is determined via map equations:
\begin{equation}
p_{N}-p_{0}=-\sum_{l=0}^{N-1}V'(\theta_{l}).
\label{app5}
\end{equation}

The last ingredient we need is the Poisson summation formula, giving the Fourier transform of a $\delta$-spectrum,
\begin{equation}
\sum_{n=-\infty}^{+\infty}\delta (y+2\pi n)=\frac{1}{2\pi}\sum_{m=-\infty}^{+\infty}\exp[imy].
\label{app6}
\end{equation}

Note that the $\delta$-function constraints $\delta(p_{i}-p_{i-1}+V'(\theta_{i-1}))$ take care of the $p$-integrals.
To perform the integration we need the \textit{Jacobi-Anger expansion}:
\begin{equation}\nonumber
e^{iz\cos\theta}=\sum_{n=-\infty}^{+\infty}i^{n}J_{n}(z)e^{in\theta},
\end{equation}
$J_{i}$ being Bessel functions of the first kind.

The $M$-th order moment at time $N$, for our specific potential, may thus be rewritten as
\begin{eqnarray}
&\langle (p_{N}-p_{0})^{M}\rangle =& \label{pMmedio} \\
\nonumber \\
&\sum_{m_{N}=-\infty}^{\infty}...\sum_{m_{1}=-\infty}^{\infty}\int_{0}^{2\pi}\frac{d\theta_{0}}{2\pi}...\int_{0}^{2\pi}\frac{d\theta_{N}}{2\pi}& \nonumber \\
 \nonumber \\
& (k\sin(\theta_{0})+k\sin(\theta_{1}-\phi)+ ... +k\sin(\theta_{N-1}-\phi))^{M} & \nonumber \\
 \nonumber\\
& \exp \left[ \right. i \sum_{r=1}^{N}m_{r}(\theta_{r}-\theta_{r-1}-(t_{r}-t_{r-1}) & \nonumber \\
 \nonumber\\
& (p_{0}+k\sin(\theta_{0})+k\sin(\theta_{1}-\phi)+ ... +k\sin(\theta_{r-1}*)))\left] \right. & \nonumber
%\label{pMmedio}
\end{eqnarray}
where $r$ is a kick index, $t_{r}$ is the time and $m_{r}$ are integers; $N$ is the last kick taken into consideration and we suppose it to be even.  $M$ is equal to $1$ for the first moment and to $2$ for the second one. \\
$\theta_{r-1}*$ is equal to $\theta_{r-1}$ if $r$ is odd and to $\theta_{r-1}-\phi$ if $r$ is even; similarly, $t_{r}-t_{r-1}$ is equal to $1-\xi$ if $r$ is even and to $\xi$ if $r$ is odd.

Since we are interested in the behavior of the map after the application of a couple of kicks, we define a new time variable: $n=N/2$.

\subsection{The current}
We want to estimate the current $\langle p-p_{0}\rangle $, i.e. equation (\ref{pMmedio}) when $M$ is $1$. The so-called quasi linear result corresponds to setting all $m_j=0$ (fully random propagator) and is null in the present case.
By taking only one $m_{j} \neq 0$ (note that it has to be $\pm 1$ in order not to have a vanishing integral) we get the first non zero contribution to the average $p$. This contribution is referred to in the literature as one kick correlation contribution; it corresponds to the lowest harmonic contribution to the Fourier series for the $\phi$ dependence of the current.
\begin{eqnarray}
& \langle p-p_{0}\rangle (n,k,\phi,\xi)^{m_{j}=\pm1}=&  \\
\nonumber\\
& -k\sin(\phi+(1-\xi)p_{0})J_{1}[(1-\xi)k] & \nonumber\\
\nonumber\\
& J_{0}[(1-\xi)k]\frac{1-(J_{0}[(1-\xi)k]^{2})^{n-1}}{1-J_{0}[(1-\xi)k]^{2}}+ & \nonumber \\
\nonumber\\
& k\sin(\phi-\xi p_{0})J_{1}[\xi k] \frac{1-(J_{0}[\xi k]^{2})^{n}}{1-J_{0}[\xi k]^{2}}  & \nonumber
\label{p_one}
\end{eqnarray}
(we have summed the terms coming from complex conjugate ($m_{j}=\pm1$) for every possible $j$-choice).
%$J_{i}$ is the $i$th order Bessel function of the first kind. 

The next frequency appears when we include the two kick correlations. This contribution is dominated by terms coming from evenly spaced kicks with coefficients of equal value, i.e.  $m_{j}=m_{j-2l}=\pm1$. In the absence of any informations about the value of $\xi k$, we take into consideration all contributions of this kind, summing over $l$ (for small $\xi k$ the dominant endowment is given by $j$ and $j-2l$ as far as possible). 
%the next neighbor terms, i.e. $m_{j}=\pm1$ and $m_{j-1}=\mp1$.
We get
%\begin{eqnarray}
%& \langle p\rangle (t,k,\phi,\xi)^{2 \; kick}=& \nonumber \\
%\nonumber\\
%& k(\sin(2\phi+(1-2\xi)p_{0})J_{1}[(1-2\xi)k] &\nonumber\\
%\nonumber\\
%&\left( \right.J_{2}[\xi k] \frac{1-(J_{0}[(2\xi-1)k]^{2})^{n}}{1-J_{0}[(2\xi-1)k]^{2}}+ & \nonumber \\
%\nonumber\\
%& J_{2}[(1-\xi) k]J_{0}[(1-2\xi) k] \frac{1-(J_{0}[(1-2\xi)k]^{2})^{n}}{1-J_{0}[(1-\xi) k]^{2}} \left. \right) &
%\label{p_two}
%\end{eqnarray}
\begin{eqnarray}
& \langle p-p_{0}\rangle (n,k,\phi,\xi)^{m_{j}=\pm1, \, m_{j-2l}=\pm1}=& \\
\nonumber\\
& k(\sin(2\phi+2(1-\xi)p_{0})J_{1}[(1-\xi)k]^{2}  J_{1}[2(1-\xi)k]& \nonumber\\
\nonumber\\
&\Big( f_{1}(J_{0}[(1-\xi)k],J_{0}[2(1-\xi)k])\Big.+&\nonumber\\
\nonumber\\
&\Big.J_{0}[2(1-\xi)k] \frac{1-(J_{0}[2(1-\xi)k]^{2})^{n-2}}{1-J_{0}[2(1-\xi)k]^{2}}\Big)+ & \nonumber \\
\nonumber\\
& k(\sin(-2\phi+2\xi p_{0})(J_{1}[\xi k]^{2}  J_{1}[2\xi k]&\nonumber\\
\nonumber\\
&\Big(f_{2}(J_{0}[\xi k],J_{0}[2\xi k])+ \frac{1-(J_{0}[2\xi k]^{2})^{n-1}}{1-J_{0}[2\xi k]^{2}}\Big) &\nonumber 
\label{p_two}
\end{eqnarray}
where $f_{1}$ and $f_{2}$ are given by
\begin{eqnarray}\nonumber
f_{1}(a,b)&=&\frac{ab}{a^{2}-b^{2}}\left(\frac{1-(a^{2})^{n-2}}{1-a^{2}}-\frac{1-(b^{2})^{n-2}}{1-b^{2}}\right) \nonumber \\
\nonumber\\
f_{2}(a,b)&=&\frac{a}{a^{2}-b^{2}}\left(\frac{1-(a^{2})^{n-1}}{1-a^{2}}-\frac{1-(b^{2})^{n-1}}{1-b^{2}}\right). \nonumber 
%\label{p_two}
\end{eqnarray}
The next  contributions would come from $m_{j}=\pm 1$ and $m_{j-(2l+1)}=\pm 1$ or  $\mp 1$, but their relevance is much smaller than the previous ones.

\begin{figure}[!htb]
\begin{center}
\includegraphics[height=6.0cm]{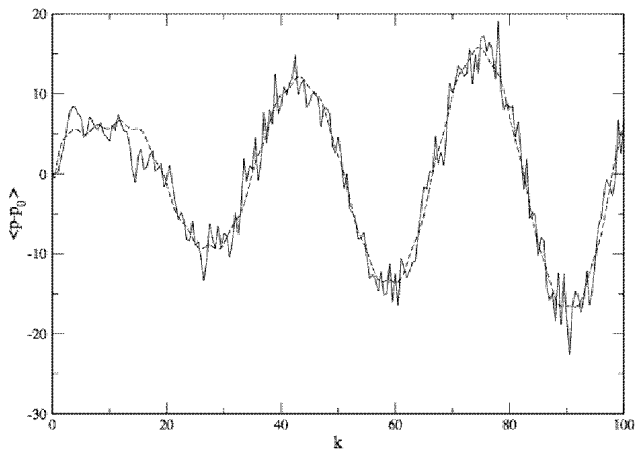}
\end{center}
\caption{\small{$\langle p-p_{0}\rangle $ after 50 couples of kicks, averaged over 100000 initial conditions, versus $k$. $\xi=0.2$, $\phi=\pi/2$, $p_{0}=0$. Full line: numerical results; dashed line: analytic estimate.}}
\label{fig10}
\end{figure}

\begin{figure}[!htb]
\begin{center}
\includegraphics[height=6.0cm]{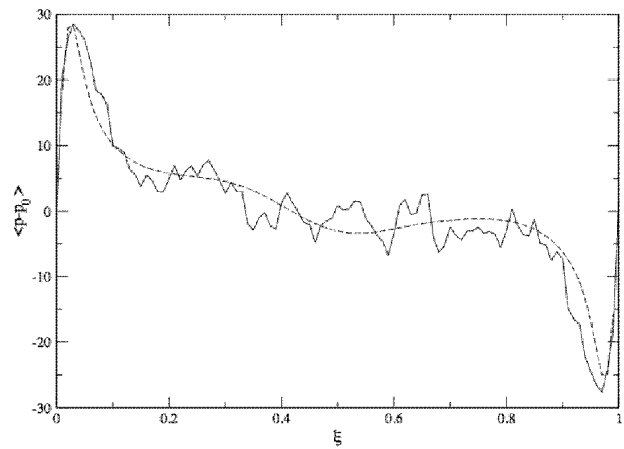}
\end{center}
\caption{\small{$\langle p-p_{0}\rangle $ after 50 couples of kicks, averaged over 100000 initial conditions, versus $\xi$. $k=9$, $\phi=\pi/2$, $p_{0}=0$. Full line: numerical results; dashed line: analytic estimate.}}
\label{fig11}
\end{figure}

\begin{figure}[!htb]
\begin{center}
\includegraphics[height=6.0cm]{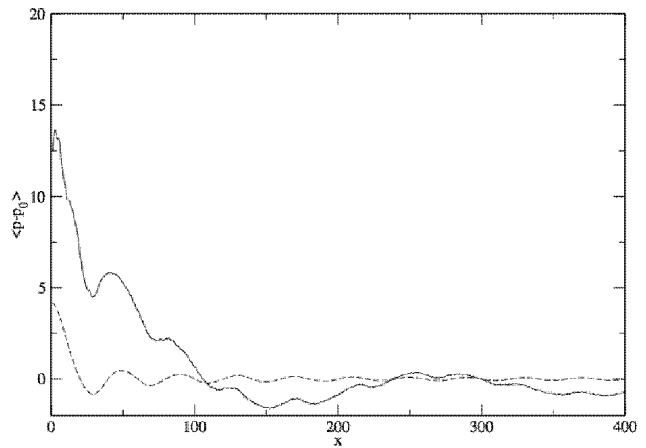}
\end{center}
\caption{\small{$\langle p-p_{0}\rangle $ after 50 couples of kicks, averaged over 100000 initial conditions, calculated over wider and wider intervals of initial conditions. $p_{0}$ is distributed in $(-x/2,x/2)$ where $x$ is the value in the $abscissa$-axis. $k=9$, $\xi=\pi/10$, $\phi=\pi/2$. Full line: numerical results; dashed line: analytic estimate.}}
\label{fig12}
\end{figure}

Now we can compare these analytic results with the numerical data (figs. (\ref{fig2}, \ref{fig10}, \ref{fig11})). The agreement is satisfactory.
%\\
%\texttt {F10--> 10}
\\
\\
%\texttt {F11--> 12}
\\
\\
In particular we remark again that by properly choosing ${\cal M}_{p_0}$ we may obtain non-zero asymptotic values of the momentum, and current reversal by a transformation on the phase shift $\phi$. If instead of a single valued initial momentum ${\cal M}_{p_0}$ we choose initial conditions distributed over the entire torus we have that the current goes to zero, both numerically and analytically. This can be done only for a rational value of $\xi$ (otherwise the torus unitary cell is not defined). For an irrational $\xi$ the current goes to zero if we calculate it over wider and wider intervals of initial conditions (see fig. (\ref{fig12})).%The absence of an elementary cell may be overcome in the irrational $\xi$ case by looking at the behavior of the average current over wider and wider intervals of initial conditions (see fig. (\ref{fig12})).
%\\
%\texttt{F12--> 16} 
\\
\\

\subsection{Diffusion}
In a similar way we can obtain analytic results for the behavior of the variance $\langle (p-p_{0})^{2}\rangle - \langle p-p_{0} \rangle ^{2}$.

The quasi linear approximation (i.e. taking all $m_{j}=0$) gives $D_{QL}=k^{2}/4$ ($D$ is the diffusion coefficient: $\langle (p-p_{0})^{2}\rangle - \langle p-p_{0} \rangle ^{2}|_{m_{j}=0}=2 D_{QL} N$).% (or, in a different notation $\frac{k^{2}}{2}$).

Then we take only one $m_{j}\neq0$; in this case we have a non zero integral by putting either $m_{j}=\pm1$ or $m_{j}=\pm2$:
\begin{eqnarray}
&\langle (p-p_{0})^{2}\rangle (n,k,\phi,\xi)^{m_{j}=\pm1}=&\\
&&\nonumber\\
&k^{2}\cos(\phi+(1-\xi)p_{0})\left( (J_{0}-J_{2})J_{0}\frac{1-(J_{0}^{2})^{n-1}}{1-J_{0}^{2}}+\right.&\nonumber\\
&&\nonumber\\
&\left.-2J_{1}^{2}\frac {1+J_{0}^{2}-(2n-1)(J_{0}^{2})^{n-1}+(2n-3)(J_{0}^{2})^{n}} {(1-J_{0}^{2})^{2}} \right)+&\nonumber\\
&&\nonumber\\
&+k^{2}\cos(\phi-\xi p_{0})\left( (J_{0}-J_{2})\frac{1-(J_{0}^{2})^{n}}{1-J_{0}^{2}}+\right.&\nonumber\\
&&\nonumber\\
&\left.-2J_{1}^{2}\left(2J_{0}\frac {1+(n-1)(J_{0}^{2})^{n}-n(J_{0}^{2})^{n-1}}{(1-J_{0}^{2})^{2}}\right) \right)&\nonumber
\label{mom2_1_2}
\end{eqnarray}
where the argument of the Bessel functions in the first part is $\left[(1-\xi)k\right]$ and in the second part $\left[\xi k\right]$.
\begin{eqnarray}
&\langle (p-p_{0})^{2}\rangle (n,k,\phi,\xi)^{m_{j}=\pm2}=&\\
&&\nonumber\\
&-\frac{k^{2}}{2}\cos(2\phi+2(1-\xi)p_{0})&\nonumber\\
&&\nonumber\\
&J_{2}[2(1-\xi)k]J_{0}[2(1-\xi)k]\frac{1-(J_{0}[2(1-\xi)k]^{2})^{n-1}}{1-J_{0}[2(1-\xi)k]^{2}}&\nonumber\\
&&\nonumber\\
&-\frac{k^{2}}{2}\cos(2\phi-2\xi p_{0})J_{2}[2\xi k]\frac{1-(J_{0}[2\xi k]^{2})^{n}}{1-J_{0}[2\xi k]^{2}}.&\nonumber
\label{mom2_1_2}
\end{eqnarray}

The point now is to classify the contribution coming from a different choice of the $m_{j}$. With two $m_{j}\neq0$ the dominant contribution comes from evenly spaced kicks with coefficients of opposite value, i.e $m_{j}=-m_{j-2l}=\pm 1$; the main endowment in this case is independent from the size of $\xi k$ and is found by choosing terms as near as possible, namely the relevant contributions come from $l=1,\,2,\,3$ .
\begin{eqnarray}
&\sum_{l=1,\,2,\,3}\langle (p-p_{0})^{2}\rangle(n,k,\phi,\xi) ^{m_{j}=\pm1, \, m_{j-2l}=\mp1}=&\\
&&\nonumber\\
&-k^{2}J_{1}[(1-\xi)k]^{2}&\nonumber\\
&&\nonumber\\
&\Big((n-2)+J_{0}[(1-\xi)k]^{2}(n-3)+J_{0}[(1-\xi)k]^{4}(n-4)\Big)&\nonumber\\
&&\nonumber\\
&-k^{2}J_{1}[\xi k]^{2}&\nonumber\\
&&\nonumber\\
&\Big((n-1)+J_{0}[\xi k]^{2}(n-2)+J_{0}[\xi k]^{4}(n-3)\Big).&\nonumber
\label{mom2_1_3}
\end{eqnarray}

\begin{figure}[!htb]
\begin{center}
\includegraphics[height=6.0cm]{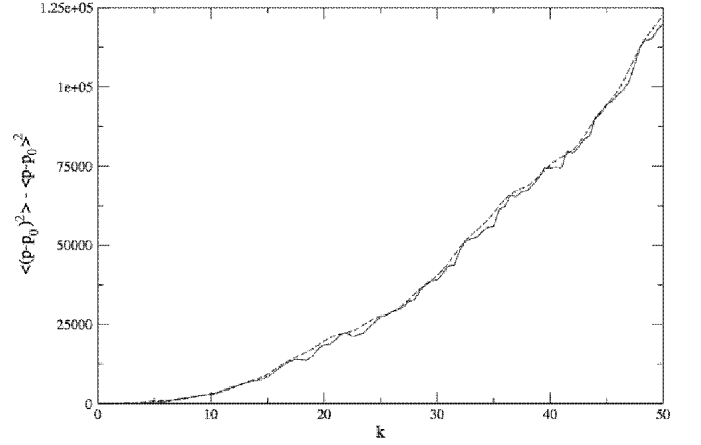}
\end{center}
\caption{\small{$\langle (p-p_{0})^{2}\rangle-\langle p-p_{0} \rangle^{2}$ after 50 couples of kicks, averaged over 100000 initial conditions, versus $k$. $\xi=0.2$, $\phi=\pi/2$, $p_{0}=0$. Full line: numerical results; dashed line: analytic estimate.}}
\label{fig13}
\end{figure}

Summing up these contributions and comparing the result with numerics, we obtain a good agreement for the behavior of $\langle (p-p_{0})^{2}\rangle - \langle p-p_{0} \rangle ^{2}$ versus the initial momentum $p_{0}$, the kick strength $k$ and the number of kicks couples $n$, see figs. (\ref{fig4}, \ref{fig5}, \ref{fig13}). About the dependence on $k$, notice that the contributions with some $m_{j}\neq 0$ only take care of the oscillations around an average value, given by the quasi linear result.
%\\
%\texttt{F13--> 19}
\\
\\

\section{Dissipation}
As we mentioned in the introduction the symmetry breaking paves the way for obtaining asymptotic non zero current, yet the linear growth in time of the variance broadens the momentum distribution, thus masking the asymmetry effect. 

A way to get a pristine ratchet behavior is to introduce dissipation \cite{copi,2harmd}, while keeping the dynamics strictly deterministic.

To this end we modify the pair of maps in the following way:
\begin{equation}
\left\{
\begin{array}{lcl}
p_{n} & = & \gamma p_{n-1} +k\sin(\theta_{n-1})\\
\\
\theta_{n} & = & \theta_{n-1}+p_{n}\xi
\end{array}
 \right.
 \label{map1d}
\end{equation}
\begin{equation}
\left\{
\begin{array}{lcl}
p_{n+1} & = &\gamma  p_{n} +k\sin(\theta_{n}-\phi)\\
\\
\theta_{n+1} & = & \theta_{n}+p_{n+1}(1-\xi)
\end{array}
 \right.
 \label{map2d}
\end{equation}
where $0\le \gamma \le 1$; if $\gamma =1$ we recover the hamiltonian system (no dissipation), while $\gamma=0$ is the overdamped case.

We can also write equations (\ref{map1d}) and (\ref{map2d}) in the form of a single map including both kicks ($(n-1)\to n$):
%
%\vspace{1cm}
\begin{equation}
\left\{
\begin{array}{lcl}
p_{n+1}&=&\gamma^{2} p_{n} +\gamma k\sin(\theta_{n})+\\
\\
&&k\sin(\theta_{n}+\xi\gamma p_{n} +\xi k\sin(\theta_{n})-\phi)\\
\\
\theta_{n+1}&=&\theta_{n}+\xi\gamma p_{n}+(1-\xi)\gamma^{2}p_{n}+\\
\\
&&(\xi+(1-\xi) \gamma)k\sin(\theta_{n})+\\
\\
&&(1-\xi) k\sin(\theta_{n}+\xi\gamma p_{n} +\xi k\sin(\theta_{n})-\phi)
\end{array}
 \right. \nonumber
 \label{map3d}
\end{equation}
%\vspace{1cm}

In this case the presence of the time asymmetry $\xi$ is not required for the ratchet effect and it can be put equal to zero or to one half; in that case the ratchet effect is purely due to dissipation. On the contrary  $\phi$ is obviously still needed in order to break the space symmetry.

\begin{figure}[!htb]
\begin{center}
\includegraphics[height=6.3cm]{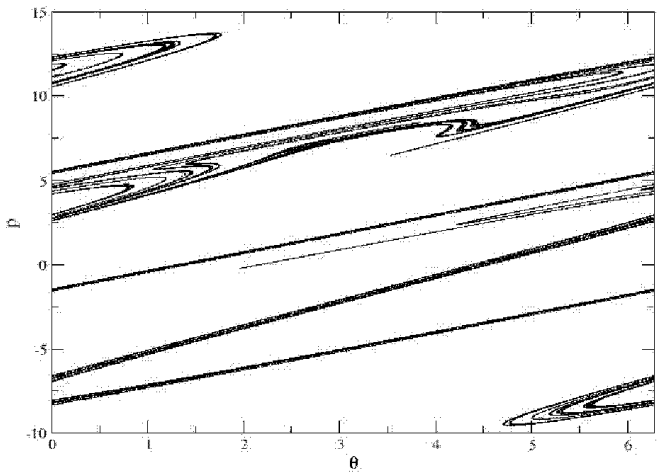}
\end{center}
\caption{\small{Phase space surface of section: $k=9$, $\phi=\pi/2$, $\xi=0.2$, $\gamma=0.4$.}}
\label{fig14}
\end{figure}

\begin{figure}[!htb]
\begin{center}
\includegraphics[height=6.0cm]{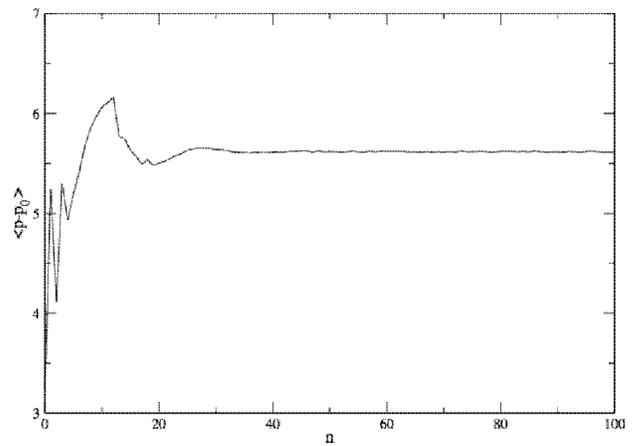}
\end{center}
\caption{\small{Average current $\langle p-p_{0}\rangle $, averaged over $5 \; 10^{6}$ initial conditions,  versus time, expressed in couples of kicks. $p_{0}=0$, $k=9$, $\xi=0.2$, $\phi=\pi/2$, $\gamma=0.4$.}}
\label{fig15}
\end{figure}

\begin{figure}[!htb]
\begin{center}
\includegraphics[height=6.0cm]{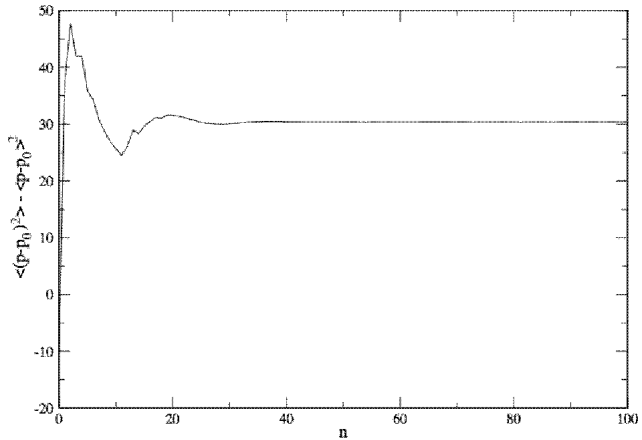}
\end{center}
\caption{\small{$\langle (p-p_{0})^{2}\rangle -\langle p-p_{0}\rangle ^{2}$, averaged over $5 \; 10^{6}$ initial conditions, versus time, expressed in couples of kicks. $p_{0}=0$, $k=9$, $\xi=0.2$, $\phi=\pi/2$, $\gamma=0.4$.}}
\label{fig16}
\end{figure}

\begin{figure}[!htb]
\begin{center}
\includegraphics[height=6.0cm]{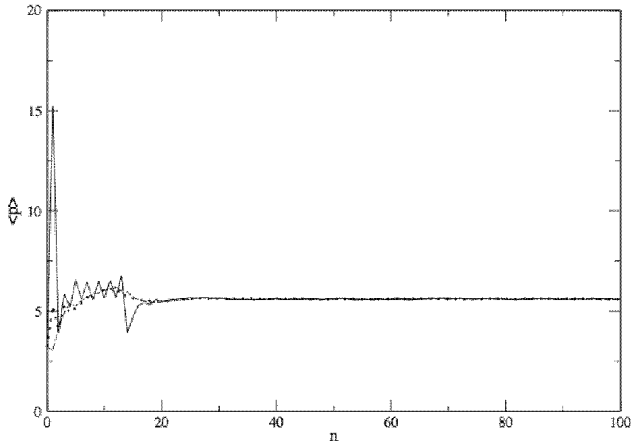}
\end{center}
\caption{\small{$\langle p \rangle $, averaged over $100000$ initial conditions, versus time, expressed in couples of kicks for different values of $p_{0}$. Dot line: $p_{0}=0$, dashed line: $p_{0}=20$, full line: $p_{0}=100$  $k=9$, $\xi=0.2$, $\phi=\pi/2$, $\gamma=0.4$.}}
\label{fig18}
\end{figure}

\begin{figure}[!htb]
\begin{center}
\includegraphics[height=6.0cm]{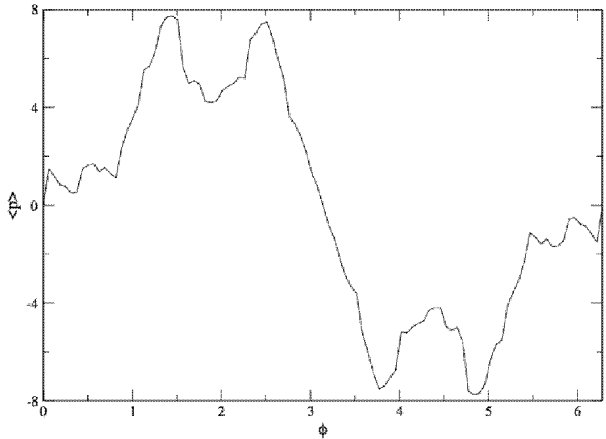}
\end{center}
\caption{\small{$\langle p \rangle $, after $50$ couples of kicks, averaged over $100000$ initial conditions, versus $\phi$. $p_{0}=0$, $k=9$, $\xi=0.2$, $\gamma=0.4$.}}
\label{fig19}
\end{figure}

\begin{figure}[!htb]
\begin{center}
\includegraphics[height=6.0cm]{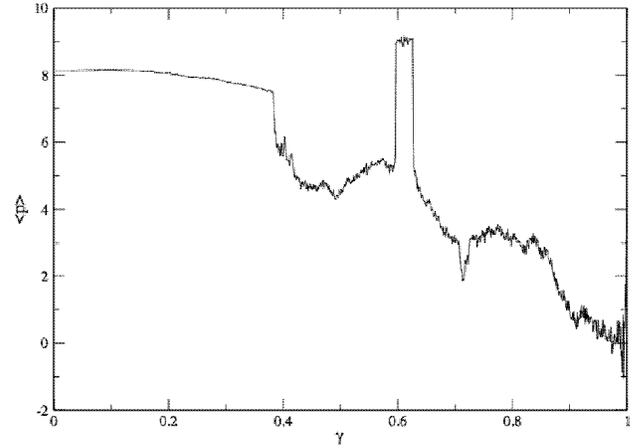}
\end{center}
\caption{\small{$\langle p \rangle $ after $500$ couples of kicks, averaged over $10000$ initial conditions, versus $\gamma$. $k=9$, $\xi=0.2$, $\phi=\pi/2$, $p_{0}=0$.}}
\label{fig21}
\end{figure}

It is known that very weak dissipation may lead to a quite complex organization of the dynamics (see \cite{Celso}), with many stable orbits with interwoven basins of attraction: deeper in the dissipative regime we typically observe, after a transient time, either an attracting periodic orbit, or a strange attractor (see fig. (\ref{fig14})).
%\\
%\texttt{F14--> 28}
\\
\\
The most interesting features in this case are the following: \textit{i.)} if space symmetry is broken the attractor is not generally symmetric in $p$ direction, thus we expect a nonvanishing asymptotic current (see fig. (\ref{fig15}))
%\\
%\texttt{F15--> 34}
\\
\\
\textit{ii.)} In contrast to the conservative case here the finite size of the attractor prevents unbounded broadening of the distribution: after some transient time the variance saturates, see fig. (\ref{fig16}).
%\\
%\texttt{F16-->  figura varianza}
\\
\\
\textit{iii.)} Motion on the attractor seems ergodic, so the dependence on ${\cal M}_{p_0}$ here disappears, and $\langle p_n \rangle$ tends to an asymptotic value which seems independent on the initial probability density (see fig. (\ref{fig18})). \textit{iv.)} Again by the $\phi \to -\phi$ we get current reversal (see fig. (\ref{fig19})). \\
In figure (\ref{fig21}) we plot the behavior of current versus $\gamma$.
%\\
%\texttt{F17--> <p> vs p0}
%\\
%\begin{figure}[!htb]
%\begin{center}
%\includegraphics[height=5.0cm]{p_p0.eps}
%\end{center}
%\caption{\small{$\langle p \rangle $, averaged over $100000$ initial conditions, versus $p_{0}$. $k=9$, $\xi=0.2$, $\phi=\pi/2$, $\gamma=0.4$.}}
%\label{fig17}
%\end{figure}
%\\
%\\
%\texttt{F18--> <p> asintotico}
\\
\\
%\\
%\texttt{F19--> 32}
\\
\\
\\
\\

We are able to give an analytic estimate for the first moment also in the dissipative case, given some quite strict conditions on parameters: $\gamma\lesssim1$ (i.e. small dissipation) and $\xi k >1$. The last condition is related to the fact that when $\gamma$ is close to $1$, the dissipative phase space is similar to the conservative one and the present approximation doesn't work if the phase space has a cellular structure.

The general formula for the analytic first moment is 
\begin{eqnarray}
&\langle p_{N}-p_{0} \rangle =& \label{pMmedioDiss} \\
\nonumber \\
&\sum_{m_{N}=-\infty}^{\infty}...\sum_{m_{1}=-\infty}^{\infty}\int_{0}^{2\pi}\frac{d\theta_{0}}{2\pi}...\int_{0}^{2\pi}\frac{d\theta_{N}}{2\pi}\Big(\Big.p_{0}(\gamma^{N}-1)+&\nonumber\\
\nonumber\\
& k\sin(\theta_{N-1}-\phi)+\gamma k\sin(\theta_{N-2})+ ... +\gamma^{N-1} k\sin(\theta_{0})\Big)\Big. & \nonumber \\
 \nonumber\\
& \exp \Big[ \Big. i \sum_{r=1}^{N}m_{r}\Big(\Big.\theta_{r}-\theta_{r-1}-(t_{r}-t_{r-1}) (p_{0}\gamma^{r}+ & \nonumber \\
 \nonumber\\
&k\sin(\theta_{r-1}^{*})+\gamma k\sin(\theta_{r-2}^{*})+ ... +\gamma^{r-1} k\sin(\theta_{0}))\Big.\Big)\Big.\Big] & \nonumber
%\label{pMmedio}
\end{eqnarray}
(notation as previously).\\
The quasi linear result (i.e. setting all $m_{j}=0$) is no more null, unless $p_{0}=0$: 
\begin{equation}
\langle p-p_{0}\rangle (n,k,\phi,\xi)^{m_{j}=0}=p_{0}(\gamma^{2n}-1).
\label{p_zero_diss}
\end{equation}
The next contribution is obtained by taking only one $m_{j}=\pm1$:
\begin{eqnarray}
& \langle p-p_{0}\rangle (n,k,\phi,\xi)^{m_{j}=\pm1} =& \\
\nonumber\\
& -k\sin(\phi+(1-\xi)p_{0})J_{1}[(1-\xi)k]J_{0}[(1-\xi)k]&\nonumber\\
\nonumber\\
&\gamma^{2n-3} \frac{ 1-((J_{0}[(1-\xi)k]/\gamma))^{2})^{n-1}}{1-(J_{0}[(1-\xi)k]/\gamma)^{2}}& \nonumber \\
\nonumber\\
& +k\sin(\phi-\xi p_{0})J_{1}[\xi k] \gamma^{2n-2}  \frac{ 1-((J_{0}[\xi k]/\gamma))^{2})^{n}}{1-(J_{0}[\xi k]/\gamma)^{2}}.& \nonumber
\label{p_one}
\end{eqnarray}
%where $f_{3}$ and $f_{4}$ are given by
%\begin{eqnarray}\nonumber
%f_{3}(k,\xi,\gamma)&=&\frac{J_{0}[(1-\xi)k]+(1-\xi)k(1-\gamma)J_{1}[(1-\xi)k]}{\gamma} \nonumber \\
%\nonumber\\
%f_{4}(k,\xi,\gamma)&=&\frac{J_{0}[\xi k]+\xi k(1-\gamma)J_{1}[\xi k]}{\gamma}.\nonumber 
%\end{eqnarray}

\begin{figure}[!htb]
\begin{center}
\includegraphics[height=6.0cm]{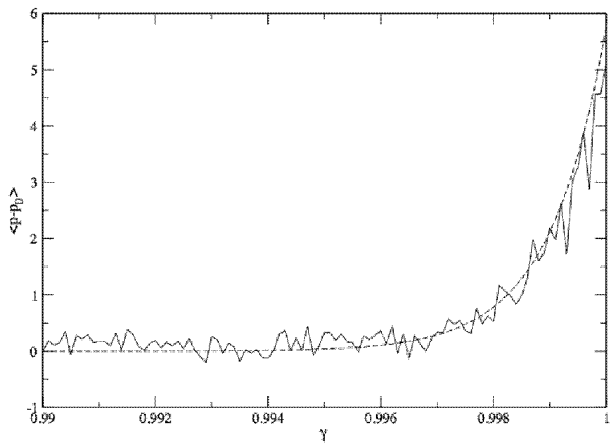}
\end{center}
\caption{\small{$\langle p-p_{0}\rangle $ after $500$ couples of kicks, averaged over $100000$ initial conditions, versus $\gamma$. $k=9$, $\xi=0.2$, $\phi=\pi/2$, $p_{0}=0$. Full line: numerical results; dashed line: analytic estimate.}}
\label{fig20}
\end{figure}

Such an estimate reasonably reproduces numerical data for a fixed value of $n$ (see fig.(\ref{fig20})), while getting meaningful asymptotic results is still an open problem.
\\
\\

\section{Conclusion}
We have considered a two dimensional area preserving map, obtained by kicking a rotator twice, with the same strength, but with a phase shift. If such a shift, and the interval between successive kicks are chosen in such a way to break relevant symmetries, then the average momentum gain may differ from zero, but the broadness of momentum distribution hides the effect in the long time limit. The dependence of transport indices on the map parameters and on the initial distribution presents interesting features, which are studied both by numerical simulations and by analytic estimates.

The most natural way to freeze (in a classical framework) the width of momentum distribution is to introduce dissipation (still keeping a strictly deterministic dynamics): typically a strange attractor arises, and transport moments localize to a finite value which does not depend upon the choice of initial probability distribution; the phase shift can then be easily tuned to get current reversal.
%%%%%%%%%%%%%%% \acknowledgments %%%%%%%%%%%%%%%%%%%%%%%%%%%%%%

This work has been partially supported by MIUR--PRIN 2005 projects %{\em 
%Order and chaos in nonlinear extended systems: coherent structures, weak 
%stochasticity and anomalous transport} 
{\em 
Transport properties of classical and quantum systems} and {\em Quantum computation with trapped particle arrays, neutral and charged}. 
We thank Gabriel Carlo for sharing his 
early work on the problem, and Giuliano Benenti for several discussions. 
%\clearpage


\begin{thebibliography}{99}
\bibitem{cosin} Similarly we can face the case far from the exact trapping momentum: $\delta p> > k^{(\xi)}/2$, getting a cosinusoidal momentum map: $p_{2}^{(\xi)}\simeq p_{0}^{(\xi)}-k^{(\xi)}\delta p\cos(\theta_{0})$.
\end{thebibliography}
\end{document}